%% file: CameraReady2027.tex
\documentclass[letterpaper]{article} 
\usepackage{aaai2027}  
\usepackage[hyphens]{url}  
\usepackage{graphicx} 
\usepackage{natbib}  
\usepackage{caption} 
\usepackage{algorithm}
\usepackage{algpseudocode}
\usepackage{amsmath}
\usepackage{amssymb}
\usepackage{amsfonts}
\usepackage{bm}
\usepackage{multirow}
\usepackage{booktabs}
\usepackage{makecell}
\usepackage{xcolor}
\usepackage{svg}
\usepackage{float}
\usepackage{placeins}
\usepackage{cuted}

\usepackage{newfloat}
\usepackage{listings}
\DeclareCaptionStyle{ruled}{labelfont=normalfont,labelsep=colon,strut=off}

\floatstyle{ruled}
\newfloat{listing}{tb}{lst}{}
\floatname{listing}{Listing}

\title{Discovering Efficient and Explainable Communication Topologies for LLM-based Multi-Agent Systems via Causal Inference}

\author{
    Junzhi Li\textsuperscript{\rm 1,\rm 2}\equalcontrib,
    Peng He\textsuperscript{\rm 3}\equalcontrib,
    Qirui Ji\textsuperscript{\rm 2}\equalcontrib,
    Wei Wang\textsuperscript{\rm 3},
    Lixiang Liu\textsuperscript{\rm 2},
    Chuxiong Sun\textsuperscript{\rm 2}\corresponding
}

\affiliations{
    \textsuperscript{\rm 1}University of Chinese Academy of Sciences\\
    \textsuperscript{\rm 2}National Key Laboratory of Space Integrated Information System,
    Institute of Software, Chinese Academy of Sciences\\
    \textsuperscript{\rm 3}Beijing University of Posts and Telecommunications\\
    lijunzhi25@mails.ucas.ac.cn,
    \{jiqirui2022, lixiang, chuxiong2016\}@iscas.ac.cn,\\
    \{hepeng123, wangwei\}@bupt.edu.cn
}

\begin{document}

\maketitle

\begin{abstract}
The performance of large language model (LLM)-based multi-agent systems (MAS) largely depends on effective communication topologies. Existing topology generation methods, however, typically learn communication topologies through black-box optimization driven solely by task-level rewards. While effective, such optimization provides little insight into why particular communication edges are selected, making it difficult to identify the critical communication subgraphs responsible for successful collaboration. To address this limitation, we propose E2-Explainer, a model-agnostic framework for providing interpretable explanations of communication topologies produced by arbitrary topology generators. Specifically, we formulate topology explanation as a causal attribution problem that identifies compact communication subgraphs supported by edge-level evidence of task preservation. We obtain this evidence with a Granger-style objective that measures how masking each communication channel changes the task outcome and the stability of the final response. The resulting budgeted subgraphs are then distilled into an amortized explainer, enabling efficient post-hoc explanation without repeated edge-level evaluations at deployment. Extensive experiments on multiple reasoning and coding benchmarks demonstrate that E2-Explainer identifies critical communication subgraphs that preserve successful collaboration. These subgraphs can also be executed directly to prune redundant communication edges, substantially reducing communication costs while maintaining competitive task performance.

\end{abstract}


\section{Introduction}
Large language model (LLM)-based agents have demonstrated remarkable capabilities across a wide range of complex tasks, including question answering~\cite{qa}, code generation~\cite{zhang2024codeagent}, and autonomous driving~\cite{jin2023surrealdriver}. However, organizing multiple LLM-based agents into a cohesive team capable of exhibiting human-like collective intelligence poses a range of new challenges~\cite{du2023improving,wu2023autogen,li2023camel,qian2023chatdev,hong2023metagpt}. Among these challenges, a fundamental one lies in designing the communication topology—that is, determining how agents should be connected and how they should transmit information during collaboration—as it directly shapes information flow, coordination effectiveness, and ultimately the collective intelligence of the system.


Early studies typically relied on manually specified and predefined topologies~\cite{du2023improving,wu2023autogen,li2023camel,qian2023chatdev,hong2023metagpt}. Although simple and effective, these topologies are largely task-agnostic, limiting their adaptability to diverse tasks and evolving collaboration demands. To improve task adaptivity and communication efficiency, recent works have increasingly explored learnable communication topology designers, which broadly fall into two paradigms. Pruning-based approaches begin with an existing collaboration workflow and remove unnecessary agents or communication links through learned gating mechanisms, structured dropout, or importance estimation~\cite{zhang2025cutthecrap,wang-etal-2025-agentdropout,li2025adaptive}. Generation-based approaches, by contrast, formulate topology design as a task-conditioned graph generation problem and synthesize collaboration topologies using autoregressive node-and-edge decoding, mixture-of-experts graph generators, or guided graph diffusion~\cite{li2026assemble,li2026ofamas,jiang2025gtd,zhang2026radar}.
\begin{figure*}[t]
    \centering
    \includegraphics[width=1\textwidth]{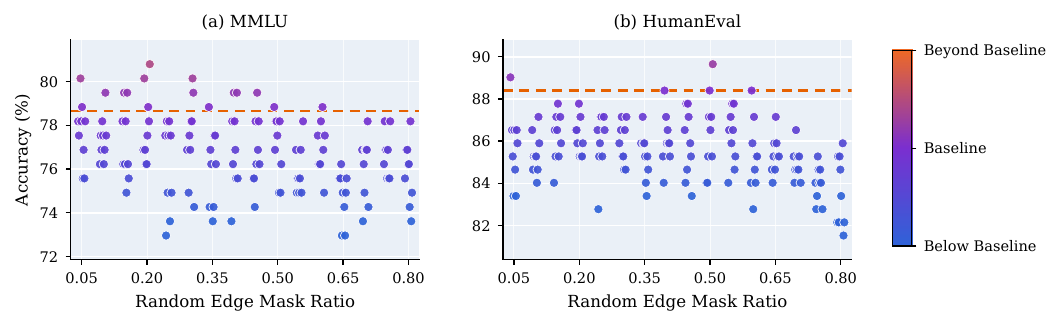}
    \caption{
    Random edge masking on G-Designer-generated communication graphs for MMLU and HumanEval.
    The edge masking ratio ranges from 0.05 to 0.80 in increments of 0.05, with ten random seeds evaluated at each ratio.
    Each point denotes the accuracy obtained with one random mask, while the dashed line marks the performance of the original unmasked G-Designer graph.
    Most random masks degrade performance, whereas only a few preserve or improve accuracy, indicating that task-preserving compact subgraphs exist but are difficult to identify through random pruning.
    }
    \label{fig:motivation_random_mask}
\end{figure*}

Despite their impressive performance, we observe that the communication topologies generated by existing methods still contain substantial redundancy. As illustrated in Figure~\ref{fig:motivation_random_mask}, removing several communication links in the topologies generated by G-Designer~\cite{pmlr-v267-zhang25cu} has little impact on the final task performance and, in some cases, even improves it. This observation suggests that the learned topologies may contain not only redundant edges but also communication patterns that are merely correlated with, rather than causally relevant to, successful collaboration. Such redundancy is difficult to discover under the black-box optimization paradigm adopted by existing methods, where topologies are optimized solely using topology-level task rewards without explicitly attributing the final outcome to individual communication decisions or substructures. Consequently, existing topology designers provide little insight into which communication substructures are truly responsible for successful collaboration. This motivates the need for a post-hoc explanation framework that can identify critical causal communication substructures, thereby revealing why a learned topology succeeds and enabling redundant communication edges to be pruned for improved efficiency.

In this work, we explicitly investigate the causal contribution of individual communication edges to the task outcomes achieved by the overall communication topology. Our formulation is inspired by the notion of Granger causality, which considers $x_i \rightarrow Y$ causal if the information provided by $x_i$ improves the prediction of $Y$~\cite{granger1969investigating,wiener1956theory}. Analogously, in a communication topology, if the information transmitted through an edge improves the cooperative performance of the MAS, that edge can be regarded as causally important to the resulting collaboration. Building upon this intuition, we extend the Granger-style criterion from individual communication edges to local communication substructures, aiming to identify compact causal subgraphs that preserve task-relevant interactions while eliminating redundant communication.

Specifically, we propose E2-Explainer, a model-agnostic post-hoc framework for identifying compact causal subgraphs from communication topologies generated by arbitrary topology designers. Given a communication topology, E2-Explainer estimates the Granger-style causal contribution of each edge by systematically masking it and measuring the resulting change in a task evaluation signal. This signal is primarily defined by the final task outcome and is supplemented with result-level semantic entropy changes to provide denser feedback under sparse rewards. The estimated edge contributions are then integrated with structural constraints to identify compact causal subgraphs that preserve task-relevant communication while eliminating redundant interactions. Since this attribution process requires repeated executions of the underlying LLM-based MAS, we further train an amortized explainer using the discovered causal subgraphs as supervision. At deployment, the explainer directly predicts a compact causal subgraph from a newly generated topology, avoiding repeated edge-masking evaluations. Extensive experiments across six reasoning and coding benchmarks demonstrate that E2-Explainer consistently improves the performance--communication cost trade-off of multiple topology designers while exhibiting strong transferability across different designers and agent scales.

Our contributions are summarized as follows:

\begin{itemize}
    \item We formulate post-hoc explanation of optimized LLM-MAS communication graphs as the identification of compact subgraphs that preserve the task behavior of the original topology, and introduce a Granger-style criterion for evaluating the contribution of communication edges.

    \item We propose E2-Explainer, which distills edge-level causal evidence obtained from offline masking evaluations into budget-specific subgraph supervision and trains an amortized explainer to directly generate compact, collaboration-preserving communication subgraphs without requiring edge-contribution evaluations at test time.

    \item We conduct extensive experiments across multiple benchmarks, topology optimizers, and agent scales, showing that E2-Explainer reduces communication costs while maintaining competitive task performance and exhibits cross-generator and cross-scale transferability.
\end{itemize}
\section{Related Work}

\paragraph{LLM-based multi-agent systems.}
LLM-based multi-agent systems organize multiple language-model agents into collaborative workflows, where agents exchange intermediate solutions, critique one another, and aggregate final decisions~\cite{du2023improving,wu2023autogen,li2023camel,qian2023chatdev,hong2023metagpt}. They build on decomposed reasoning, self-consistency, and multi-perspective deliberation to improve LLM reliability~\cite{wei2022chain,wang2023selfconsistency,kojima2022large}, and have been summarized in recent surveys on agent workflows and applications~\cite{guo2024large,li2024survey,chen2024survey}. Our work focuses on their communication structure rather than new roles or prompting strategies.

\paragraph{Communication topology optimization.}
A growing line of work studies how to design efficient communication topologies for LLM-MAS. Recent methods can be roughly grouped into generative topology methods, which construct task-adaptive collaboration graphs, such as G-Designer and ARG-Designer~\cite{pmlr-v267-zhang25cu,li2026assemble}, and pruning or selection methods, which remove unnecessary links, agents, or graph components from an existing workflow, such as AgentPrune, AgentDropout, Cut the Crap, and adaptive graph pruning methods~\cite{zhang2025cutthecrap,wang-etal-2025-agentdropout,li2025adaptive,cang2026graphgrpo}. These methods show that topology strongly affects the performance--cost trade-off. E2-Explainer is complementary: it treats generated topologies as candidate graphs and learns to refine them into smaller task-preserving communication subgraphs.

\paragraph{Causality in machine learning.}
Causality provides a principled framework for reasoning about interventions, counterfactuals, and invariant mechanisms~\cite{pearl2009causality,peters2017elements,scholkopf2021toward}. In machine learning, causal ideas have been used to analyze model behavior under controlled perturbations and to identify factors that are responsible for predictions~\cite{chattopadhyay2019neural,schwab2019cxplain}. Granger causality characterizes causal informativeness by whether access to one variable improves the prediction of another~\cite{granger1969investigating,granger1980testing,bressler2011wiener}. In graph learning, this intuition has been connected to explanation methods that identify compact graph structures responsible for preserving or changing model behavior, such as GNNExplainer, PGExplainer, counterfactual explainers, and GEM~\cite{ying2019gnnexplainer,luo2020parameterized,pmlr-v151-lucic22a,pmlr-v139-lin21d}. Inspired by these ideas, we adapt Granger-style causal reasoning from differentiable graph models to frozen LLM-MAS communication graphs. Unlike prior graph explanation methods that primarily produce explanations for inspection, E2-Explainer distills offline edge-masking evidence into an amortized explainer that directly outputs budgeted task-preserving communication subgraphs at deployment.

\section{Problem Formulation}

\subsection{LLM-MAS Communication Graphs}

We consider an LLM-based multi-agent system with agents $V=\{a_1,\ldots,a_n\}$. Given query $x$, a topology generator $\mathcal{G}_{\phi}$ produces a directed communication graph
\begin{equation*}
G=\mathcal{G}_{\phi}(x)=(V,E),
\end{equation*}
where $e=(a_i,a_j)\in E$ delivers the output of agent $a_i$ to agent $a_j$. Executing $G$ yields a final response $y(x,G)$ and a dataset-specific task score $Q(x,G)$, such as exact-match accuracy, execution correctness, or an LLM-judged score.

\subsection{Post-hoc Communication Explanation}

Given a generated graph $G$, we seek a compact and executable subgraph
\begin{equation*}
H=(V_H,E_H),\quad H\subseteq G,
\end{equation*}
that preserves its task behavior. The retained edges explain the communication paths supporting the original collaboration, while executing $H$ may reduce communication overhead. We denote the realized online cost by $C(x,H)$, measured using actual execution tokens. Because removing an edge changes downstream prompts and messages, $C(x,H)$ need not vary monotonically with the number of retained edges or nodes.

\subsection{Objective and Deployment Constraint}

Let $b=(\rho_E,\rho_V)$ denote a subgraph budget, where $\rho_E$ and $\rho_V$ specify the desired edge-pruning and node-pruning ratios, and let $\mathcal{H}_b(G)$ be the corresponding family of compact subgraphs. A valid explanation under budget $b$ should satisfy
\begin{equation}
H_b^{\star}\in\mathcal{H}_b(G),\qquad
Q(x,H_b^{\star})\approx Q(x,G).
\label{eq:posthoc_refinement_objective}
\end{equation}
The budget determines compactness, while the task-behavior condition defines faithfulness. We report $C(x,H_b^{\star})$ as measured efficiency rather than assuming that structural sparsity guarantees a fixed token reduction. Because validating candidate subgraphs requires repeated LLM-MAS executions and may require task feedback, we learn an amortized explainer $F_{\theta}$:
\begin{equation}
\widehat{H}_b
=
F_{\theta}(x,G,b),
\quad
\widehat{H}_b\subseteq G.
\end{equation}

\begin{figure*}[t]
    \centering
    \includegraphics[width=\textwidth]{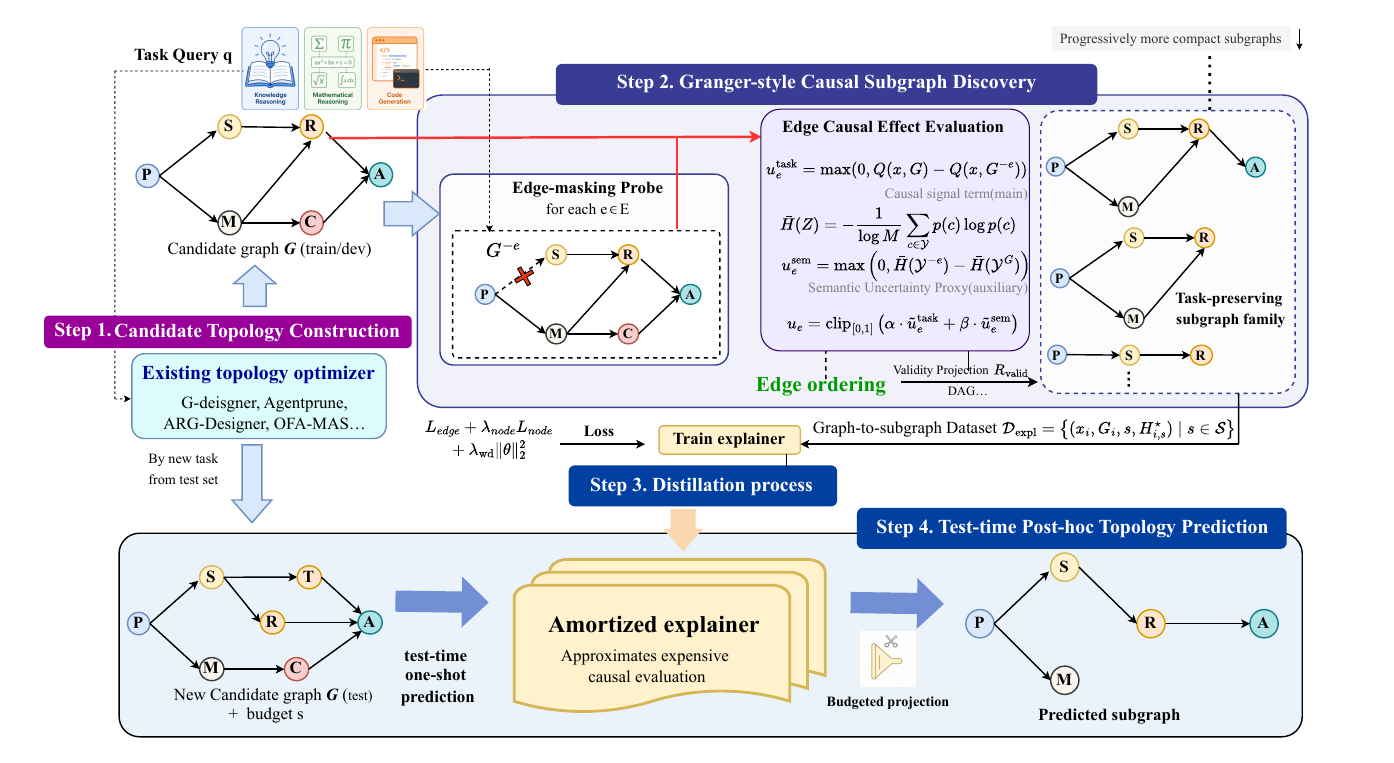}
    \caption{Overview of E2-Explainer. Edge masking estimates preservation utilities from task-score and semantic-entropy changes, which yield budgeted subgraphs for supervision. The amortized explainer then predicts a task-preserving subgraph for an unseen topology in one forward pass.}
    \label{fig:method}
\end{figure*}
\section{E2-Explainer}

As illustrated in Figure~\ref{fig:method}, E2-Explainer is a model-agnostic post-hoc framework for identifying compact causal subgraphs from communication topologies generated by arbitrary topology designers. Given a communication topology, E2-Explainer estimates the Granger-style causal contribution of each edge by systematically masking it and measuring the resulting change in a task evaluation signal. This signal is primarily derived from the final task outcome and further incorporates topology-level semantic entropy changes to provide denser feedback when the task outcome is sparse or coarse-grained. The estimated causal contributions are then combined with budget constraints and structural requirements to construct executable compact subgraphs that preserve task-relevant communication patterns. Since this attribution process requires repeated edge-masking evaluations of the underlying LLM-based MAS, we use the discovered subgraphs as supervision to train an amortized explainer, which can directly predict a budget-specific subgraph for a newly generated topology at test time.

\subsection{Edge-Level Granger-Style Causal Attribution}

To estimate the Granger-style causal contribution of each communication edge under the original topology, we perform single-edge removal interventions while keeping all other factors fixed. For a query $x$, graph $G$, and edge $e=(u,v)$, we preserve the same agents, prompts, decoding configurations, and all other communication channels, while intervening on $e$ by blocking the message transmitted from source agent $u$ to target agent $v$. Following the Granger criterion, we quantify the contribution of $e$ by measuring the change in the evaluation signal before and after this intervention.

\paragraph{Task-level causal contribution.}
For each active edge $e\in E$ of the generated graph $G=(V,E)$, we construct the intervened graph $G^{-e}=G\setminus\{e\}$ and re-execute the frozen LLM-MAS. Let $Q(x,G)$ and $Q(x,G^{-e})$ denote the task scores obtained before and after intervening on $e$, respectively. We define the task-level causal contribution of $e$ as
\begin{equation}
    \Delta^{\mathrm{task}}_e
    =
    Q(x,G)-Q(x,G^{-e}).
    \label{eq:task_effect}
\end{equation}
A positive value indicates that retaining edge $e$ improves the task outcome relative to removing it, whereas a negative value suggests that the information transmitted through $e$ may be detrimental to the task outcome. Since our explanation focuses on communication that supports the original collaboration, we retain only the nonnegative preservation contribution:
\begin{equation}
    u_e^{\mathrm{task}}
    =
    \max
    \left(
    0,\Delta_e^{\mathrm{task}}
    \right).
    \label{eq:task_utility}
\end{equation}
A larger value provides stronger evidence that edge $e$ contributes to preserving the task behavior of the candidate graph, while zero indicates no positive preservation evidence under the observed task metric.


\paragraph{Auxiliary semantic signal.}
The final task outcome can be too coarse to distinguish edge masks that receive the same task score. We therefore use result-level semantic entropy as an auxiliary signal to characterize the stability of the final response~\cite{kuhn2023semantic,farquhar2024detecting}. For the response set $\mathcal{Y}$ obtained from $M$ stochastic executions, we group semantically equivalent responses into classes $\mathcal{C}(\mathcal{Y})$ and compute
\begin{equation}
\bar{H}(\mathcal{Y})
=
-
\sum_{c\in\mathcal{C}(\mathcal{Y})}
p(c)\log p(c),
\label{eq:semantic_entropy}
\end{equation}
where $p(c)$ is the empirical frequency of class $c$ among the $M$ executions. We use $M=5$ independently sampled executions for both the original graph and each edge-masked graph. The auxiliary semantic effect of edge $e$ is defined as
\begin{equation}
u^{\mathrm{sem}}_e
=
\max\left(
0,\,
\bar{H}(\mathcal{Y}^{G^{-e}})
-
\bar{H}(\mathcal{Y}^{G})
\right).
\label{eq:semantic_proxy}
\end{equation}
A positive value indicates that masking $e$ increases the uncertainty of the final response, providing additional evidence that the edge contributes to preserving the behavior of the original graph.


\paragraph{Unified edge effect.}
We combine the task-level causal effect with the auxiliary semantic stability signal. The resulting edge utility is
\begin{equation}
    u_e
    =
    \operatorname{clip}_{[0,1]}
    \left(
    \alpha \cdot \widetilde{u}^{\mathrm{task}}_e
    +
    \beta \cdot \widetilde{u}^{\mathrm{sem}}_e
    \right),
    \label{eq:edge_utility}
\end{equation}
where $\widetilde{\cdot}$ denotes normalization within the current graph or calibration batch, and $\alpha,\beta\geq 0$ control the two signals. The utilities induce a preservation ordering over the edges of $G$:
\begin{equation}
    \pi_G=(e_{(1)},e_{(2)},\ldots,e_{(|E|)}),
    \quad
    u_{e_{(1)}}\geq u_{e_{(2)}}\geq \cdots \geq u_{e_{(|E|)}}.
    \label{eq:preservation_ordering}
\end{equation}
This ordering connects edge-level Granger-style attribution to budgeted subgraph construction.

\subsection{Causal Subgraph Extraction}
The edge-level causal contributions obtained above provide a principled basis for extracting compact communication subgraphs. Given a target budget $b$, E2-Explainer selects edges according to their preservation utilities and constructs a subgraph that retains the most task-relevant information flows while satisfying the structural constraints inherited from the original workflow. This budget-aware extraction process achieves a desirable trade-off between communication sparsity and task preservation without requiring exhaustive subgraph enumeration. Specifically, the budget determines the number of retained edges, while the causal ordering determines which communication paths are preserved.

\paragraph{Budget-Aware Subgraph Selection.}
The budget $b=(\rho_E,\rho_V)$ specifies the desired retention ratios of edges and optional nodes. For each graph, these ratios are converted into retention counts $k_E(b)$ and $k_V(b)$ according to the graph size. We first select the most important edges according to the preservation ordering obtained from edge attribution:
\begin{equation}
    E^{\star}_{b}
    =
    \operatorname{TopK}
    \left(
    \pi_G,
    k_E(b)
    \right).
    \label{eq:prefix_edges}
\end{equation}
The node budget is subsequently enforced jointly with the structural projection described below to obtain an executable communication subgraph.

\paragraph{Executability-Constrained Projection.}
To transform the selected high-utility edges into an executable communication topology, we apply an executability-constrained projection:
\begin{equation}
    H^{\star}_{b}
    =
    \mathcal{R}_{\mathrm{valid}}(G,E^{\star}_{b},k_V(b)).
    \label{eq:valid_projection}
\end{equation}
The projection $\mathcal{R}_{\mathrm{valid}}$ only selects edges from the original edge set $E$ and never introduces new links. Therefore, when the original topology $G$ is a directed acyclic graph, the extracted subgraph naturally preserves acyclicity. Specifically, the projection retains mandatory input and final-aggregation agents, removes communication links incident to excluded optional agents, and discards optional agents that no longer participate in valid directed communication paths. When the node budget is active, optional agents are prioritized according to the aggregated preservation utilities of their incident retained edges. The resulting $H_b^{\star}$ therefore satisfies both the specified budget constraints and the execution requirements of the underlying LLM-MAS. 

\paragraph{Budget-Specific Explanation Family.}
Applying Eq.~\eqref{eq:valid_projection} to the budget set $\mathcal{B}$ yields an explanation family
\begin{equation}
    \mathcal{F}^{\star}(x,G)
    =
    \{H^{\star}_{b}: b\in\mathcal{B}\}.
    \label{eq:explanation_family}
\end{equation}
Each member provides an explanation of the same candidate topology under a different compactness requirement. Collectively, these subgraphs characterize different trade-offs between communication efficiency and task preservation. As the retention budget increases, the extracted subgraphs progressively incorporate lower-ranked communication edges to recover additional information flows.


\subsection{Amortized Subgraph Explainer}
The causal attribution and structural projection procedures described above provide a reliable mechanism for identifying compact, task-preserving subgraphs from arbitrary communication topologies. However, directly applying this procedure to a newly generated topology at test time is computationally expensive, as it requires repeated edge-masking interventions and multiple executions of the underlying LLM-MAS to estimate edge contributions. To overcome this limitation, we amortize the causal extraction process by collecting extracted subgraphs from generated topologies as graph-to-subgraph supervision. Specifically, these causally grounded subgraphs serve as explanation targets that teach an explainer to predict which communication paths should be preserved under different compactness budgets. After training, the amortized explainer can directly generate budget-specific causal subgraphs for unseen topologies without additional causal evaluations. 


\paragraph{Graph-to-subgraph supervision.}
For each query $x_i$ and candidate graph $G_i$, the extraction procedure produces one target subgraph $H^{\star}_{i,b}$ for every budget $b\in\mathcal{B}$. These targets form the training set
\begin{equation}
    \mathcal{D}_{\mathrm{expl}}
    =
    \left\{
    (x_i,G_i,b,H^{\star}_{i,b})
    \mid
    b\in\mathcal{B}
    \right\}.
    \label{eq:subgraph_dataset}
\end{equation}
Thus, the explanation family provides supervision rather than an input to the explainer. For a fixed graph, the targets share the same causal preservation ordering but correspond to different pruning budgets. Learning from the complete family therefore teaches the explainer both which communication elements are important and how their retention changes with the requested compactness, allowing one model to serve multiple budgets.


\paragraph{Explainer mapping.}
We train a parameterized explainer $F_{\theta}$ to approximate the graph-to-subgraph mapping induced by the causal extraction procedure:
\begin{equation}
    \widehat{H}_{i,b}
    =
    F_{\theta}(x_i,G_i,b),
    \qquad
    \widehat{H}_{i,b}\subseteq G_i.
    \label{eq:explainer_mapping}
\end{equation}
Given a query, a candidate graph, and a pruning budget, $F_{\theta}$ internally predicts edge and node retention probabilities and decodes them into the corresponding budget-specific subgraph. The explainer is independent of the topology generation mechanism, with architectural details deferred to the Appendix.


\paragraph{Training objective.}
Let $A^{\star}_{i,b,e}$ indicate whether edge $e$ belongs to
$H^{\star}_{i,b}$, and let $R^{\star}_{i,b,v}$ indicate whether
node $v$ is retained. The explainer predicts the corresponding
probabilities $\widehat{A}_{i,b,e}$ and $\widehat{R}_{i,b,v}$,
which are collected into the edge- and node-retention vectors
$\widehat{\mathbf{A}}_{i,b}$ and $\widehat{\mathbf{R}}_{i,b}$,
respectively. We optimize
\begin{equation}
\begin{aligned}
    \mathcal{L}_{\mathrm{edge}}
    &=
    \sum_{(x_i,G_i,b,H^{\star}_{i,b})\in\mathcal{D}_{\mathrm{expl}}}
    \frac{1}{|E_i|}
    \sum_{e\in E_i}
    \mathrm{BCE}
    \left(
    A^{\star}_{i,b,e},
    \widehat{A}_{i,b,e}
    \right),\\
    \mathcal{L}_{\mathrm{node}}
    &=
    \sum_{(x_i,G_i,b,H^{\star}_{i,b})\in\mathcal{D}_{\mathrm{expl}}}
    \frac{1}{|V_i|}
    \sum_{v\in V_i}
    \mathrm{BCE}
    \left(
    R^{\star}_{i,b,v},
    \widehat{R}_{i,b,v}
    \right),
\end{aligned}
\label{eq:subgraph_reconstruction_loss}
\end{equation}
with the final objective
\begin{equation}
    \mathcal{L}
    =
    \mathcal{L}_{\mathrm{edge}}
    +
    \lambda_{\mathrm{node}}
    \mathcal{L}_{\mathrm{node}}
    +
    \lambda_{\mathrm{wd}}\|\theta\|_2^2.
    \label{eq:final_subgraph_loss}
\end{equation}
\begin{table*}[t]
\centering
\small
\setlength{\tabcolsep}{3.8pt}
\renewcommand{\arraystretch}{1.08}

\resizebox{\textwidth}{!}{
\begin{tabular}{llcccccc!{\vrule width 0.4pt}c}
\toprule
Method
& Metric
& MMLU
& GSM8K
& MultiArith
& SVAMP
& AQuA
& HumanEval
& Avg. \\
\midrule

Vanilla
& Acc.
& 72.78
& 88.15
& 96.67
& 93.40
& 82.68
& 81.37
& 85.84 \\

CoT
& Acc.
& 76.81
& 91.81
& 97.67
& 93.60
& 83.07
& 87.58
& 88.42 \\

SC(CoT)
& Acc.
& 77.34
& 92.48
& 97.67
& 93.83
& 84.13
& 87.08
& 88.76 \\

\midrule

ARG-Designer
& Acc.
& 76.68
& 89.18
& 97.81
& 93.57
& 83.95
& 86.78
& 88.00 \\

\multirow{2}{*}{ARG-Designer + E2-Explainer}
& Acc.
& 77.99{\scriptsize $\uparrow$1.31}
& 86.94{\scriptsize $\downarrow$2.24}
& 97.17{\scriptsize $\downarrow$0.64}
& 94.67{\scriptsize $\uparrow$1.10}
& 84.58{\scriptsize $\uparrow$0.63}
& 89.26{\scriptsize $\uparrow$2.48}
& 88.44{\scriptsize $\uparrow$0.44} \\
& Token $\Delta$
& -12.1\%
& -15.5\%
& -23.3\%
& -21.1\%
& -29.8\%
& -14.3\%
& -20.1\% \\

\midrule

OFA-MAS
& Acc.
& 79.91
& 91.01
& 98.21
& 93.97
& 84.11
& 89.26
& 89.41 \\

\multirow{2}{*}{OFA-MAS + E2-Explainer}
& Acc.
& \textbf{81.87}{\scriptsize $\uparrow$1.96}
& 90.40{\scriptsize $\downarrow$0.61}
& 97.41{\scriptsize $\downarrow$0.80}
& 92.65{\scriptsize $\downarrow$1.32}
& 86.67{\scriptsize $\uparrow$2.56}
& 90.62{\scriptsize $\uparrow$1.36}
& 89.94{\scriptsize $\uparrow$0.53} \\
& Token $\Delta$
& -15.1\%
& -27.2\%
& -25.5\%
& -23.8\%
& -17.8\%
& -26.8\%
& -25.6\% \\

\midrule

AgentPrune
& Acc.
& 78.43
& 92.78
& \textbf{98.33}
& 94.90
& 85.04
& 87.58
& 89.51 \\

\multirow{2}{*}{AgentPrune + E2-Explainer}
& Acc.
& 79.08{\scriptsize $\uparrow$0.65}
& \textbf{94.09}{\scriptsize $\uparrow$1.31}
& \textbf{98.33}{\scriptsize $\pm$0.00}
& 94.20{\scriptsize $\downarrow$0.70}
& 86.61{\scriptsize $\uparrow$1.57}
& 88.20{\scriptsize $\uparrow$0.62}
& 90.09{\scriptsize $\uparrow$0.58} \\
& Token $\Delta$
& -14.7\%
& -26.7\%
& -27.3\%
& -22.9\%
& -38.1\%
& -10.6\%
& -24.9\% \\

\midrule

G-Designer
& Acc.
& 78.65
& 93.05
& 98.13
& \textbf{95.00}
& 84.52
& 88.40
& 89.63 \\

\multirow{2}{*}{G-Designer + E2-Explainer}
& Acc.
& 79.74{\scriptsize $\uparrow$1.09}
& 93.77{\scriptsize $\uparrow$0.72}
& 98.17{\scriptsize $\uparrow$0.04}
& 94.60{\scriptsize $\downarrow$0.40}
& \textbf{87.30}{\scriptsize $\uparrow$2.78}
& \textbf{90.69}{\scriptsize $\uparrow$2.29}
& 90.71{\scriptsize $\uparrow$1.09} \\
& Token $\Delta$
& -31.5\%
& -44.0\%
& -28.0\%
& -20.1\%
& -37.5\%
& -19.3\%
& -25.1\% \\

\bottomrule
\end{tabular}
}
\caption{
Main results with Qwen3-8B as the backbone LLM.
Boldface indicates the best accuracy for each dataset.
For E2-Explainer variants, Acc. arrows indicate changes relative to the corresponding candidate generator, while the Avg. column reports average accuracy and weighted token reduction across all datasets.
}
\label{tab:main_results}
\end{table*}
\paragraph{Subgraph prediction.}
The predicted probabilities are first converted into budget-compatible candidate sets:
\begin{equation}
\begin{aligned}
    \widehat{E}_{i,b}
    &=
    \operatorname{TopK}
    \left(
    \widehat{\mathbf{A}}_{i,b},
    k_E(b)
    \right),\\
    \widehat{V}_{i,b}
    &=
    \operatorname{TopK}
    \left(
    \widehat{\mathbf{R}}_{i,b},
    k_V(b)
    \right).
\end{aligned}
\label{eq:predicted_elements}
\end{equation}
Accordingly, the final subgraph is obtained as
\begin{equation}
    \widehat{H}_{i,b}
    =
    F_{\theta}(x_i,G_i,b)
    =
    \mathcal{R}_{\mathrm{valid}}
    \left(
    G_i,
    \widehat{E}_{i,b},
    \widehat{V}_{i,b}
    \right),
    \label{eq:predicted_subgraph}
\end{equation}
where mandatory nodes are retained by $\mathcal{R}_{\mathrm{valid}}$. At deployment, this process requires neither task labels nor repeated edge interventions.


\section*{Experiment}
\subsection{Experimental Setup}

\paragraph{Datasets and backbone.}
We evaluate on six benchmarks: AQuA, GSM8K, MultiArith, and SVAMP for mathematical reasoning~\citep{ling2017program,cobbe2021gsm8k,roy2015solving,patel2021nlp}, MMLU for knowledge-intensive reasoning~\citep{hendrycks2021measuring}, and HumanEval for code generation~\citep{chen2021humaneval}. All methods use Qwen3-8B as the backbone LLM~\citep{yang2025qwen3}, with Vanilla, CoT~\citep{wei2022chain}, and SC(CoT)~\citep{wang2023selfconsistency} as single-agent baselines.

\paragraph{Candidate generators and calibration.}
We train a single explainer using only G-Designer-generated candidate graphs. For each dataset, 40 calibration inputs are sampled from the training split without overlap with the test split, and edge-masking probes construct graph-to-subgraph supervision. The frozen explainer is then applied without retraining to graphs produced by ARG-Designer, OFA-MAS, AgentPrune, and G-Designer. 

\paragraph{Subgraph setting and metrics.}
The main comparison uses $E25+N20$ as the default subgraph-size setting, while other settings are examined in the appendix. We report task accuracy and actual online token usage; changes for E2-Explainer are computed relative to the corresponding candidate generator.

\subsection{Results and Analysis}
\paragraph{Main comparison.}

Table~\ref{tab:main_results} evaluates E2-Explainer as a post-hoc module for four representative topology optimizers. Across all four optimizers, E2-Explainer reduces weighted token usage by 20.1\%--25.6\%, while preserving or improving their average accuracy. Specifically, it improves the average accuracy of ARG-Designer, OFA-MAS, AgentPrune, and G-Designer by 0.44, 0.53, 0.58, and 1.09 points, respectively. G-Designer+E2-Explainer achieves the highest average accuracy of 90.71 while consuming 25.1\% fewer tokens than the original G-Designer. These results directly support the primary objective of E2-Explainer: reducing communication overhead without sacrificing the collaborative effectiveness of an already optimized topology.

The fine-grained results show that the reduction in communication cost is more consistent than the change in task accuracy. Across the 24 optimizer--dataset combinations, E2-Explainer reduces token usage in every case, with dataset-level reductions ranging from 10.6\% to 44.0\%. Meanwhile, accuracy improves in 16 combinations and remains unchanged in one. Improvements are particularly consistent on MMLU, AQuA, and HumanEval, where all four topology optimizers benefit from post-processing. Results on GSM8K, MultiArith, and SVAMP are more mixed, indicating that the amount of removable communication depends on both the task and the initial candidate topology. Nevertheless, the universal token reduction and predominantly preserved or improved accuracy show that compact task-preserving subgraphs can often be recovered from optimized communication graphs. The additional accuracy gains further suggest that some candidate topologies still retain redundant or distracting communication whose removal can benefit answer quality.
%

A key observation is that the improvement transfers beyond the generator used for calibration. The explainer is trained only on G-Designer graphs but applied without retraining to ARG-Designer, OFA-MAS, and AgentPrune. These unseen generators have different construction biases, covering autoregressive graph generation, mixture-of-experts graph generation, and pruning-based topology optimization. Nevertheless, E2-Explainer improves their average accuracy by 0.44--0.58 points and reduces weighted online cost by 20.1\%--25.6\%. Since these generators are unseen during calibration, the transfer is unlikely to come from memorizing G-Designer-specific adjacency patterns. Instead, the explainer appears to capture reusable cues about which communication paths are necessary under a given task and graph context. 

\begin{table}[t]
\centering
\small
\setlength{\tabcolsep}{5pt}
\begin{tabular}{lcccc}
\toprule
\multirow{2}{*}{Setting} & \multicolumn{2}{c}{MMLU} & \multicolumn{2}{c}{HumanEval} \\
\cmidrule(lr){2-3} \cmidrule(lr){4-5}
& Acc. & Tok. $\downarrow$ & Acc. & Tok. $\downarrow$ \\
\midrule
Original (6-agent) & 79.43 & -- & 89.26 & -- \\
E2-Explainer (5$\rightarrow$6-agent) & \textbf{81.70} & 10.16\% & \textbf{90.08} & 19.38\% \\
\bottomrule
\end{tabular}
\caption{Generalization across agent scales on G-Designer-generated communication graphs. The explainer is trained on 5-agent graphs and evaluated on 6-agent graphs. Token reductions are measured against the original 6-agent topology.}
\label{tab:scale_generalization}
\end{table}

\paragraph{Generalization across agent scales.}
Table~\ref{tab:scale_generalization} further tests cross-scale generalization by training the explainer on 5-agent G-Designer graphs and directly evaluating it on 6-agent graphs. Without retraining, E2-Explainer improves MMLU accuracy from 79.43 to 81.70 and HumanEval accuracy from 89.26 to 90.08, while also reducing token usage by 10.16\% and 19.38\%, respectively. These results suggest that the learned explainer transfers beyond the graph size observed during training and can still identify compact task-preserving subgraphs under moderate changes in agent scale.

\begin{table}[t]
\centering
\small
\setlength{\tabcolsep}{5.0pt}
\renewcommand{\arraystretch}{1.08}

\begin{tabular}{lcccc}
\toprule
\multirow{2}{*}{Method}
& \multicolumn{2}{c}{MMLU}
& \multicolumn{2}{c}{HumanEval} \\
\cmidrule(lr){2-3}\cmidrule(lr){4-5}
& Acc.
& Token $\Delta$
& Acc.
& Token $\Delta$ \\
\midrule
w/o Causal
& 75.82
& -11.16\%
& 86.88
& -5.05\% \\
w/o Sem. Entropy
& 77.78
& -23.42\%
& 87.50
& -14.32\% \\
E2-Explainer
& \textbf{79.74}
& -31.50\%
& \textbf{90.69}
& -19.25\% \\
\bottomrule
\end{tabular}
\caption{
Component ablation of E2-Explainer on G-Designer-generated communication graphs for MMLU and HumanEval.
}
\label{tab:component_ablation}
\end{table}

\paragraph{Component ablation.}
Table~\ref{tab:component_ablation} evaluates the roles of the causal and auxiliary semantic-entropy signals. Removing the causal signal leads to a larger degradation, reducing accuracy from 79.74 to 75.82 on MMLU and from 90.69 to 86.88 on HumanEval. The causal-only variant, namely \textit{w/o Sem. Entropy}, performs better than the semantic-entropy-only variant, suggesting that task-level causal effects provide the primary preservation-aware signal for edge valuation. However, causal-only still falls short of the full model. This indicates that the result-level semantic entropy proxy complements causal supervision by densifying the refinement signal when final-task feedback is sparse.

\begin{table}[!ht]
\centering
\footnotesize
\setlength{\tabcolsep}{2.2pt}
\renewcommand{\arraystretch}{1.02}

\begin{tabular}{lrrrr}
\toprule
\multirow{2}{*}{Source}
& \multicolumn{2}{c}{MMLU}
& \multicolumn{2}{c}{HumanEval} \\
\cmidrule(lr){2-3}\cmidrule(lr){4-5}
& Acc. & Token $\Delta$ & Acc. & Token $\Delta$ \\
\midrule
\multicolumn{5}{l}{\textit{Target: G-Designer}} \\
Original
& 78.65 & -- & 88.40 & -- \\
G-Designer
& 79.74 & -31.50\% & \textbf{90.69} & -19.25\% \\
AgentPrune
& 79.08 & -14.43\% & 88.75 & -15.40\% \\
Mixed
& \textbf{80.18} & -24.55\% & 89.30 & -20.61\% \\
\midrule
\multicolumn{5}{l}{\textit{Target: AgentPrune}} \\
Original
& 78.43 & -- & 87.58 & -- \\
G-Designer
& 79.08 & -14.70\% & 88.20 & -10.60\% \\
AgentPrune
& \textbf{81.70} & -24.56\% & \textbf{90.07} & -19.18\% \\
Mixed
& 79.08 & -26.69\% & 88.82 & -13.43\% \\
\bottomrule
\end{tabular}
\caption{
Effect of calibration sources under $E25+N20$.
\textit{Original} denotes unrefined target graphs, and \textit{Mixed} combines G-Designer and AgentPrune calibration graphs.
Boldface marks the best refined accuracy.
}
\label{tab:calibration_source}
\end{table}

\paragraph{Effect of Calibration Sources.}
Table~\ref{tab:calibration_source} compares explainers trained on G-Designer graphs, AgentPrune graphs, or their mixture and evaluated on both target optimizers. All six calibration--target combinations improve accuracy while reducing token usage, providing evidence of bidirectional cross-optimizer transfer. Source matching remains beneficial: AgentPrune-only calibration achieves the highest accuracy on both datasets for AgentPrune targets, while G-Designer-only performs best on HumanEval for G-Designer targets. Mixed-source calibration obtains the best G-Designer result on MMLU and improves both target optimizers, but does not uniformly outperform single-source calibration. These results suggest that E2-Explainer learns transferable communication patterns while still benefiting from structural characteristics specific to the target optimizer.
\section{Conclusion}

We present E2-Explainer, a post-hoc framework for identifying compact and collaboration-preserving communication subgraphs from optimized LLM-MAS topologies. E2-Explainer distills offline edge-level evaluations into an amortized explainer, enabling efficient test-time generation of compact communication subgraphs that preserve collaborative task performance without requiring ground-truth answers or repeated edge-level evaluations. Experiments across multiple topology optimizers show that it reduces communication costs while maintaining task performance, and transfers across graph sources and agent scales.

\bibliography{aaai2027}
\newpage

\appendix

\input{supplementary}


\end{document}

%% file: supplementary.tex
\section{Implementation Details}
\label{app:implementation}

\subsection{Explainer Input and Output}
\label{app:explainer_io}

The amortized explainer learns the budget-conditioned mapping
\begin{equation}
    F_{\theta}:(x,G,b)\mapsto \widehat{H}_{b},
    \qquad \widehat{H}_{b}\subseteq G,
\end{equation}
where $x$ is the task query, $G=(V,E)$ is a candidate communication graph, and $b=(\rho_E,\rho_V)$ specifies the edge- and node-pruning ratios. All inputs are available before executing the predicted subgraph; the explainer does not require ground-truth answers, repeated edge interventions, or internal parameters of the topology generator.

For each edge $e=(u,v)\in E$ and node $v\in V$, the network predicts an edge-retention probability $\widehat{A}_{b,e}\in[0,1]$ and a node-retention probability $\widehat{R}_{b,v}\in[0,1]$. These probabilities are intermediate outputs rather than the final explanation. They are converted into budget-compatible candidate sets and passed through the validity projection to obtain the executable subgraph $\widehat{H}_{b}$.

The explainer therefore learns a budget-conditioned graph-to-subgraph mapping. The same candidate graph can yield different explanations under different pruning budgets, while the supervision is defined by the edge and node membership of the corresponding compact target subgraph $H_b^{\star}$.


\subsection{Budget-Conditioned Explainer Architecture}
\label{app:explainer_architecture}

\paragraph{Input representations.}
Let $\mathbf{q}_x$ denote a fixed embedding of query $x$, and let $\mathbf{r}_v$ denote a fixed embedding of the role or system prompt assigned to agent $v$. Each node is also associated with a structural feature vector $\mathbf{t}_v$, containing graph-local information available from the candidate topology, such as normalized in-degree and out-degree and topological position. A shared node encoder produces
\begin{equation}
    \mathbf{h}_v
    =
    \phi_{\mathrm{node}}
    \left(
    [\mathbf{q}_x\Vert\mathbf{r}_v\Vert\mathbf{t}_v]
    \right),
    \qquad
    \mathbf{g}_G
    =
    \frac{1}{|V|}\sum_{v\in V}\mathbf{h}_v,
    \label{eq:appendix_node_encoding}
\end{equation}
where $\Vert$ denotes concatenation and $\mathbf{g}_G$ is a permutation-invariant graph-level summary. The pruning ratios are encoded as
\begin{equation}
    \mathbf{z}_b=\phi_b([\rho_E,\rho_V]).
    \label{eq:appendix_budget_encoding}
\end{equation}

\paragraph{Edge and node prediction heads.}
For an active edge $e=(u,v)$, let $\mathbf{s}_e$ collect edge-local structural features, including the edge type and normalized structural statistics of its endpoints. The shared edge head predicts
\begin{equation}
    \widehat{A}_{b,e}
    =
    \sigma\!\left(
    \phi_{\mathrm{edge}}
    ([\mathbf{h}_u\Vert\mathbf{h}_v\Vert\mathbf{g}_G\Vert\mathbf{z}_b\Vert\mathbf{s}_e])
    \right).
    \label{eq:appendix_edge_head}
\end{equation}
The node head predicts
\begin{equation}
    \widehat{R}_{b,v}
    =
    \sigma\!\left(
    \phi_{\mathrm{retain}}
    ([\mathbf{h}_v\Vert\mathbf{g}_G\Vert\mathbf{z}_b])
    \right).
    \label{eq:appendix_node_head}
\end{equation}
All node and edge instances share the same encoders and prediction heads. Consequently, the number of trainable parameters does not depend on the numbers of agents or communication links, which allows the explainer to process graphs of different sizes.

The modules $\phi_{\mathrm{node}}$, $\phi_b$, $\phi_{\mathrm{edge}}$, and $\phi_{\mathrm{retain}}$ are implemented as lightweight multilayer perceptrons with parameters shared across graph elements. Fixed task and role representations are computed once and reused during explainer training and inference.


\subsection{Graph-to-Subgraph Supervision and Decoding}
\label{app:supervision_decoding}

For each calibration pair $(x_i,G_i)$, causal subgraph construction produces one target $H_{i,b}^{\star}$ for every budget $b\in\mathcal{B}$. The resulting training set is
\begin{equation}
    \mathcal{D}_{\mathrm{expl}}
    =
    \{(x_i,G_i,b,H_{i,b}^{\star})\mid b\in\mathcal{B}\}.
\end{equation}
Each target subgraph defines binary edge- and node-retention labels
\begin{equation}
    A_{i,b,e}^{\star}=\mathbb{I}[e\in E(H_{i,b}^{\star})],
    \qquad
    R_{i,b,v}^{\star}=\mathbb{I}[v\in V(H_{i,b}^{\star})].
\end{equation}
The explainer is optimized with the edge and node binary-cross-entropy losses defined in the main paper. Because one candidate graph contributes one target for each budget, the model learns both which communication elements should be retained and how the selected subgraph changes with the requested pruning ratios.

At inference time, the predicted probabilities are converted into candidate sets according to the graph-specific retention counts:
\begin{equation}
    \widehat{E}_b
    =\operatorname{TopK}(\widehat{\mathbf{A}}_b,k_E(b)),
    \qquad
    \widehat{V}_b
    =\operatorname{TopK}(\widehat{\mathbf{R}}_b,k_V(b)).
\end{equation}
The final explanation is
\begin{equation}
    \widehat{H}_b
    =
    \mathcal{R}_{\mathrm{valid}}(G,\widehat{E}_b,\widehat{V}_b),
\end{equation}
where $\mathcal{R}_{\mathrm{valid}}$ enforces the requested node budget together with a minimum-node constraint, removes every edge incident to an excluded node, and retains only feasible directed links between selected agents. The projection ranks feasible elements using the predicted retention scores and selects only nodes and edges already present in $G$, so it never introduces a new communication link.

\subsection{Causal Supervision Construction Details}
\label{app:causal_details}

\paragraph{Repeated execution and task score.}
For every active edge $e$, we independently execute both the original graph $G$ and the intervened graph $G^{-e}$ for $M=5$ runs. The task score is the mean outcome over these executions:
\begin{equation}
Q(x,G)
=
\frac{1}{M}
\sum_{m=1}^{M}Q^{(m)}(x,G).
\label{eq:appendix_mean_task_score}
\end{equation}
The original graph is re-executed for each evaluated edge rather than sharing one response set across all interventions. All executions use a decoding temperature of $0.2$. The original and intervened runs are sampled independently and do not use paired random seeds. The same response sets are used to compute both the mean task outcome and the auxiliary semantic signal.

\paragraph{Result-level semantic signal.}
Semantic entropy is computed from the final system responses, rather than from the local outputs of the two agents incident to an edge. This choice aligns the auxiliary signal with the system-level quantity whose causal contribution is being estimated. The $M$ final responses are grouped into semantic equivalence classes using the same equivalence procedure for the original and intervened graphs, after which the empirical class entropy is computed as defined in the main paper.

\paragraph{Score normalization.}
The mean task outcome and its non-negative intervention difference already lie in $[0,1]$. We therefore retain the task contribution on its original scale and normalize the semantic contribution by the maximum empirical entropy of $M$ samples:
\begin{equation}
\widetilde{u}^{\mathrm{task}}_e
=
u^{\mathrm{task}}_e,
\qquad
\widetilde{u}^{\mathrm{sem}}_e
=
\min\left(
1,\frac{u^{\mathrm{sem}}_e}{\log M}
\right).
\label{eq:appendix_score_normalization}
\end{equation}
The combined preservation utility is then computed with $(\alpha,\beta)=(0.8,0.2)$.

\paragraph{Pruning counts and validity projection.}
The budget $b=(\rho_E,\rho_V)$ specifies removal ratios. For a candidate graph $G=(V,E)$, the numbers of removed elements are
\begin{equation}
r_E(b)=\left\lceil \rho_E |E| \right\rceil,
\qquad
r_V(b)=\left\lceil \rho_V |V| \right\rceil,
\label{eq:appendix_pruning_counts}
\end{equation}
with retention counts $k_E(b)=|E|-r_E(b)$ and $k_V(b)=|V|-r_V(b)$. Thus, E25 denotes removing $25\%$ of the active edges and N20 denotes removing $20\%$ of the active nodes. After budgeted selection, the validity projection applies the selected node mask, removes incident edges, and retains the highest-ranked feasible links subject to the edge budget and the minimum-node constraint.

\paragraph{Scope of the causal attribution.}
The intervention procedure estimates a first-order conditional contribution for each edge in the context of the remaining full topology. Following the Granger principle, it measures whether excluding one information channel reduces the mean outcome or the stability of the final response while all other channels remain available. We do not enumerate multi-edge coalitions, whose cost grows combinatorially with graph size. Instead, these conditional edge contributions provide a tractable signal for constructing graph-level supervision, while the end-to-end evaluation of the predicted subgraph measures the aggregate outcome after multiple communication elements are removed.


\subsection{Implementation Settings}
\label{app:hyperparameters}

Table~\ref{tab:appendix_hyperparameters} summarizes the main settings used for causal supervision construction and explainer training. We train one dataset-specific explainer for each benchmark. The six explainers use the same architecture, optimization configuration, and budget set, but each is trained from the 40 calibration queries of its corresponding dataset. The same frozen Qwen3-8B backbone is used for all methods.

\begin{table}[H]
\centering
\small
\setlength{\tabcolsep}{3.5pt}
\renewcommand{\arraystretch}{1.05}
\begin{tabular}{@{}ll@{}}
\toprule
\textbf{Item} & \textbf{Setting} \\
\midrule
Semantic samples $M$ & 5 \\
Task/semantic weights $(\alpha,\beta)$ & $(0.8,0.2)$ \\
Calibration queries per dataset & 40 \\
Training budgets $\mathcal{B}$ &
\begin{tabular}[c]{@{}l@{}}
E25+N0, E50+N0, E75+N0,\\
E25+N20, E50+N20, E25+N40
\end{tabular} \\
Shared hidden dimension & 128 \\
Dropout & 0.15 \\
Optimizer & AdamW \\
Learning rate & $1\times10^{-3}$ \\
Weight decay & $1\times10^{-4}$ \\
Batch size / epochs & 64 / 200 \\
Backbone LLM & Qwen3-8B \\
Decoding temperature & 0.2 \\
Reported evaluation runs & 3 \\
Hardware & 4$\times$ NVIDIA GeForce RTX 4090 GPUs \\
Inference framework & PyTorch and vLLM \\
Token accounting & Online prompt + completion tokens \\
\bottomrule
\end{tabular}
\caption{Main implementation settings. Dataset-specific explainers share the same architecture and optimization configuration.}
\label{tab:appendix_hyperparameters}
\end{table}


\subsection{Dataset and Evaluation Details}
\label{app:dataset_details}


We evaluate E2-Explainer on six reasoning and code-generation benchmarks using the data scale and evaluation protocol adopted by G-Designer and AgentPrune. For datasets with a dedicated training split, 40 calibration queries are selected from that split and are not used as evaluation queries. For HumanEval, we follow the AgentPrune data split: 40 problems are used for calibration and the remaining 124 problems are used for evaluation. G-Designer, AgentPrune, OFA-MAS, ARG-Designer, and all variants equipped with E2-Explainer are evaluated on exactly the same query set for each benchmark.

\begin{table}[H]
\centering
\scriptsize
\setlength{\tabcolsep}{2.7pt}
\renewcommand{\arraystretch}{1.05}
\begin{tabular}{@{}llrrl@{}}
\toprule
\textbf{Dataset} & \textbf{Task} & \textbf{Calib.} & \textbf{Eval.} & \textbf{Metric} \\
\midrule
MMLU       & Knowledge MCQ   & 40 & 153   & Accuracy \\
GSM8K      & Math reasoning  & 40 & 1,319 & Exact match \\
MultiArith & Math reasoning  & 40 & 600   & Exact match \\
SVAMP      & Math reasoning  & 40 & 1,000 & Exact match \\
AQuA       & Algebra MCQ     & 40 & 254   & Accuracy \\
HumanEval  & Code generation & 40 & 124   & Pass@1 \\
\bottomrule
\end{tabular}
\caption{Dataset statistics and evaluation protocols. HumanEval uses 40 calibration problems and the remaining 124 problems for evaluation. All methods use identical evaluation queries.}
\label{tab:appendix_datasets}
\end{table}

\paragraph{MMLU.}
MMLU evaluates knowledge-intensive reasoning across subjects from STEM, the humanities, and the social sciences. Each query contains a question and four candidate answers. We evaluate on the 153-query validation subset used by G-Designer and AgentPrune and report multiple-choice accuracy.

\paragraph{GSM8K, MultiArith, and SVAMP.}
These benchmarks evaluate multi-step arithmetic reasoning over natural-language word problems. Their evaluation sets contain 1,319, 600, and 1,000 queries, respectively. Performance is measured by exact match after extracting and normalizing the final numerical answer.

\paragraph{AQuA.}
AQuA contains algebraic word problems with five candidate answers. We evaluate on 254 queries and report multiple-choice accuracy based on the final selected option.

\paragraph{HumanEval.}
HumanEval evaluates function-level Python code generation from a function signature and natural-language specification. Following AgentPrune, 40 problems are used for calibration and the remaining 124 problems are used for evaluation. Pass@1 is computed by executing the generated implementation against the associated unit tests.

\paragraph{Answer extraction and scoring.}
We follow the answer-processing conventions of the released G-Designer evaluation implementation, which are also used by AgentPrune for the overlapping benchmarks. For MMLU, the evaluator locates the final answer statement and extracts the option label from \texttt{A}--\texttt{D}; AQuA is processed analogously with labels \texttt{A}--\texttt{E}. For GSM8K, MultiArith, and SVAMP, the parser first checks for an explicit final-answer statement or a \texttt{\textbackslash boxed\{\}} expression and otherwise uses the last numerical expression in the response. Commas, surrounding spaces, trailing punctuation, and superficial LaTeX formatting are removed before exact comparison with the normalized reference value. For HumanEval, Markdown code fences are removed and the generated Python function is executed against the provided unit tests; a query is counted as correct only when all tests pass within the execution timeout. The same extraction and scoring pipeline is used for every topology generator and its corresponding E2-Explainer variant.

\subsection{Additional Evaluation Protocols}
\label{app:evaluation_protocols}

\paragraph{Controlled candidate-graph comparison.}
For each query, a topology generator and its E2-Explainer variant start from exactly the same candidate graph and use the same agent roles, prompts, decoding temperature, and evaluation procedure. The reported accuracy and token results are averaged over three independent runs.

\paragraph{Token accounting.}
The online communication cost includes the prompt and completion tokens of all retained agent calls and the final aggregation call during execution of the candidate or predicted subgraph. It excludes the offline edge interventions, semantic-entropy sampling, topology-optimizer training, explainer training, and the lightweight explainer forward pass.

\paragraph{Causal ablations.}
Both ablations reconstruct the target graph-to-subgraph supervision and retrain the explainer. ``w/o Causal'' uses only the semantic signal by setting $(\alpha,\beta)=(0,1)$, whereas ``w/o Sem. Entropy'' uses only the task-level causal contribution by setting $(\alpha,\beta)=(1,0)$.

\paragraph{Calibration-source analysis.}
The G-Designer-only and AgentPrune-only variants each use 40 candidate graph instances generated from the corresponding 40 calibration queries. In the mixed setting, the same 40 queries are processed by both topology generators, yielding 80 candidate graph instances in total.

\paragraph{Agent-scale generalization.}
For MMLU and HumanEval, the scale-generalization experiment trains the explainer on 40 five-agent calibration graphs and directly evaluates it on six-agent graphs generated separately under a six-role G-Designer configuration. The six-agent roles are reconstructed following the role-design logic of G-Designer. The E25+N20 budget and evaluation queries are identical to those used in the main comparison.

\paragraph{Meaning of model-agnostic.}
In this work, model-agnostic specifically denotes independence from the candidate topology generator. E2-Explainer does not access generator parameters or generator-specific optimization states; it operates on the candidate graph together with the task and agent-role information.



\begin{strip}
\section{Additional Experimental Results}
\label{app:additional_results}

\subsection{Results across Different Pruning Budgets}
\label{app:budget_results}

\centering
\scriptsize
\setlength{\tabcolsep}{3.0pt}
\renewcommand{\arraystretch}{1.08}
\resizebox{\textwidth}{!}{%
\begin{tabular}{llrrrrrrr}
\toprule
Budget & Metric & MMLU & GSM8K & MultiArith & SVAMP & AQuA & HumanEval & Avg./Total \\
\midrule
Original & Acc. & 78.65 & 93.05 & 98.13 & 95.00 & 84.52 & 88.40 & 89.63 \\
 & Tokens & 985,486 & 1,549,496 & 3,068,504 & 11,377,630 & 1,340,006 & 725,946 & 19,047,068 \\
\midrule
E25+N0 & Acc. & 78.43{\scriptsize$\downarrow$0.22} & 93.39{\scriptsize$\uparrow$0.34} & 98.00{\scriptsize$\downarrow$0.13} & 94.60{\scriptsize$\downarrow$0.40} & 87.10{\scriptsize$\uparrow$2.58} & 88.75{\scriptsize$\uparrow$0.35} & 90.05{\scriptsize$\uparrow$0.42} \\
 & Tokens & 985,289{\scriptsize$\downarrow$0.02\%} & 1,472,951{\scriptsize$\downarrow$4.94\%} & 2,919,068{\scriptsize$\downarrow$4.87\%} & 11,325,293{\scriptsize$\downarrow$0.46\%} & 1,315,752{\scriptsize$\downarrow$1.81\%} & 696,908{\scriptsize$\downarrow$4.00\%} & 18,715,261{\scriptsize$\downarrow$1.74\%} \\
\midrule
E50+N0 & Acc. & 79.08{\scriptsize$\uparrow$0.43} & 91.95{\scriptsize$\downarrow$1.10} & 98.17{\scriptsize$\uparrow$0.04} & 94.30{\scriptsize$\downarrow$0.70} & 84.52{\scriptsize$\pm$0.00} & 90.62{\scriptsize$\uparrow$2.22} & 89.77{\scriptsize$\uparrow$0.15} \\
 & Tokens & 972,970{\scriptsize$\downarrow$1.27\%} & 1,496,193{\scriptsize$\downarrow$3.44\%} & 2,906,180{\scriptsize$\downarrow$5.29\%} & 11,300,262{\scriptsize$\downarrow$0.68\%} & 1,305,434{\scriptsize$\downarrow$2.58\%} & 642,172{\scriptsize$\downarrow$11.54\%} & 18,623,211{\scriptsize$\downarrow$2.23\%} \\
\midrule
E75+N0 & Acc. & 77.78{\scriptsize$\downarrow$0.87} & 92.71{\scriptsize$\downarrow$0.34} & 97.83{\scriptsize$\downarrow$0.30} & 93.80{\scriptsize$\downarrow$1.20} & 89.29{\scriptsize$\uparrow$4.77} & 89.38{\scriptsize$\uparrow$0.98} & 90.13{\scriptsize$\uparrow$0.51} \\
 & Tokens & 960,849{\scriptsize$\downarrow$2.50\%} & 1,428,945{\scriptsize$\downarrow$7.78\%} & 2,889,917{\scriptsize$\downarrow$5.82\%} & 10,812,162{\scriptsize$\downarrow$4.97\%} & 1,232,270{\scriptsize$\downarrow$8.04\%} & 718,396{\scriptsize$\downarrow$1.04\%} & 18,042,539{\scriptsize$\downarrow$5.27\%} \\
\midrule
\textbf{E25+N20} & Acc. & 79.74{\scriptsize$\uparrow$1.09} & 93.77{\scriptsize$\uparrow$0.72} & 98.17{\scriptsize$\uparrow$0.04} & 94.60{\scriptsize$\downarrow$0.40} & 87.30{\scriptsize$\uparrow$2.78} & 90.69{\scriptsize$\uparrow$2.29} & 90.71{\scriptsize$\uparrow$1.09} \\
 & Tokens & 675,452{\scriptsize$\downarrow$31.46\%} & 867,098{\scriptsize$\downarrow$44.04\%} & 2,208,095{\scriptsize$\downarrow$28.04\%} & 9,091,864{\scriptsize$\downarrow$20.09\%} & 837,906{\scriptsize$\downarrow$37.47\%} & 585,838{\scriptsize$\downarrow$19.30\%} & 14,266,253{\scriptsize$\downarrow$25.10\%} \\
\midrule
E50+N20 & Acc. & 76.47{\scriptsize$\downarrow$2.18} & 92.86{\scriptsize$\downarrow$0.19} & 97.83{\scriptsize$\downarrow$0.30} & 94.30{\scriptsize$\downarrow$0.70} & 85.71{\scriptsize$\uparrow$1.19} & 86.88{\scriptsize$\downarrow$1.52} & 89.01{\scriptsize$\downarrow$0.62} \\
 & Tokens & 774,789{\scriptsize$\downarrow$21.38\%} & 1,103,706{\scriptsize$\downarrow$28.77\%} & 2,174,956{\scriptsize$\downarrow$29.12\%} & 9,004,256{\scriptsize$\downarrow$20.86\%} & 889,764{\scriptsize$\downarrow$33.60\%} & 653,860{\scriptsize$\downarrow$9.93\%} & 14,601,331{\scriptsize$\downarrow$23.34\%} \\
\midrule
E25+N40 & Acc. & 79.64{\scriptsize$\uparrow$0.99} & 92.55{\scriptsize$\downarrow$0.50} & 98.00{\scriptsize$\downarrow$0.13} & 94.10{\scriptsize$\downarrow$0.90} & 86.11{\scriptsize$\uparrow$1.59} & 81.88{\scriptsize$\downarrow$6.52} & 88.71{\scriptsize$\downarrow$0.91} \\
 & Tokens & 643,030{\scriptsize$\downarrow$34.75\%} & 948,911{\scriptsize$\downarrow$38.76\%} & 1,714,373{\scriptsize$\downarrow$44.13\%} & 6,562,617{\scriptsize$\downarrow$42.32\%} & 575,265{\scriptsize$\downarrow$57.07\%} & 483,916{\scriptsize$\downarrow$33.34\%} & 10,928,112{\scriptsize$\downarrow$42.63\%} \\
\bottomrule
\end{tabular}%
}
\captionof{table}{Performance of E2-Explainer on G-Designer candidate graphs under different pruning budgets. Accuracy changes are measured against the unpruned G-Designer graph. Token reductions are reported per dataset, and the final column gives macro-average accuracy and total token usage across all six datasets. E25+N20 is the default setting used in the main paper.}
\label{tab:multi_budget_results}
\end{strip}

Table~\ref{tab:multi_budget_results} provides a more complete view of the accuracy--efficiency trade-off controlled by the edge- and node-pruning budgets. Edge-only pruning largely preserves overall task performance, with the three settings changing average accuracy by only $+0.42$, $+0.15$, and $+0.51$ percentage points. However, even E75+N0 reduces weighted token usage by only 5.27\%. This shows that structural sparsity alone does not translate into proportional execution savings when all agents remain active: each retained agent still performs its own reasoning, and the lengths of the remaining messages can change after the topology is modified. At the same time, the gains on AQuA and HumanEval indicate that removing low-contribution channels can suppress redundant or misleading information rather than merely reducing cost.

Introducing node pruning produces substantially larger token reductions, but the benefit depends strongly on the budget. E25+N20 provides the most favorable overall trade-off, improving average accuracy from 89.63 to 90.71 while reducing weighted token usage by 25.10\%. It preserves or improves performance on five of the six benchmarks, including gains of 1.09, 2.78, and 2.29 percentage points on MMLU, AQuA, and HumanEval, respectively. Increasing the node-pruning ratio to 40\% further reduces token usage by 42.63\%, but decreases average accuracy to 88.71, largely because HumanEval drops by 6.52 percentage points. This result suggests that aggressive node removal is particularly risky for code generation, where different roles may provide complementary implementation and verification signals. E50+N20 also removes more edges than E25+N20 but achieves both lower accuracy and a smaller token reduction. The realized cost is therefore jointly determined by node selection, edge selection, structural projection, and the lengths of the remaining messages, rather than by the nominal sparsity ratio alone.

We perform the complete budget sweep only on G-Designer candidate graphs because its purpose is to characterize budget sensitivity and select one common operating point, rather than tune a different ratio for every topology generator. After selecting E25+N20, we keep the budget fixed when transferring E2-Explainer to AgentPrune, OFA-MAS, and ARG-Designer. This isolates cross-generator generalization from generator-specific budget tuning and evaluates all topology optimizers under the same compression requirement. Selecting a separate budget for each generator on the evaluation set would introduce method-specific tuning advantages and make the comparison less controlled.


\subsection{Qualitative Case Studies and Execution Traces}
\label{app:qualitative_cases}

We examine four E25N20 examples covering all possible transitions between
correct and incorrect predictions. Table~\ref{tab:case_overview} summarizes
their outcomes and online token usage. In every example, the original
candidate graph contains six active spatial masks, whereas E25N20 retains
two active spatial masks after removing one agent. The executable acyclic
trace contains three same-round communication edges before refinement and
one afterward. Active masks describe the selected candidate topology, while
the trace records the directed messages that are actually executed after
acyclic projection.


\begin{table*}[t]
\centering
\scriptsize
\setlength{\tabcolsep}{4pt}
\renewcommand{\arraystretch}{1.10}
\begin{tabular}{@{}llcccc@{}}
\toprule
\textbf{Outcome transition} & \textbf{Dataset / index} &
\textbf{Original} & \textbf{E25N20} &
\textbf{Online tokens} & \textbf{Token change} \\
\midrule
Correct $\rightarrow$ Correct & MultiArith / 419 & 49 & 49 & 6,383 $\rightarrow$ 4,567 & $-28.5\%$ \\
Wrong $\rightarrow$ Wrong & MMLU / 62 & A & A & 6,314 $\rightarrow$ 5,550 & $-12.1\%$ \\
Correct $\rightarrow$ Wrong & MMLU / 151 & B & C & 11,811 $\rightarrow$ 10,216 & $-13.5\%$ \\
Wrong $\rightarrow$ Correct & SVAMP / 415 & 47 & 3 & 6,059 $\rightarrow$ 4,965 & $-18.1\%$ \\
\bottomrule
\end{tabular}
\caption{Overview of the four qualitative cases. Online tokens include the
prompt and completion tokens of the agent calls and final aggregation call
for the corresponding query. Per-query totals are obtained from the
sequential usage records within each evaluation scope.}
\label{tab:case_overview}
\end{table*}

The original and E25N20 outputs are independent executions with decoding
temperature $0.2$. Accordingly, the cases illustrate observed changes in
information propagation and aggregation under different topologies, rather
than paired single-sample proofs that one removed element alone causes the
change in correctness.


\paragraph{Case 1: Correct $\rightarrow$ Correct with lower token usage.}
The MultiArith query asks how much longer a painter needs to finish 12 rooms
when each room requires 7 hours and 5 rooms have already been painted. The
reference answer is $(12-5)\times 7=49$. The original trace is
$\texttt{4tCg}\!\rightarrow\!\texttt{3LK5}$,
$\texttt{4tCg}\!\rightarrow\!\texttt{3Ssf}$, and
$\texttt{3LK5}\!\rightarrow\!\texttt{3Ssf}$. All four agents return 49.
E25N20 removes \texttt{4tCg} (Mathematical Analyst), leaving only
$\texttt{3LK5}\!\rightarrow\!\texttt{3Ssf}$; the three retained agents
and the final node still return 49.


\begin{table*}[t]
\centering
\scriptsize
\setlength{\tabcolsep}{4pt}
\renewcommand{\arraystretch}{1.12}
\begin{tabular}{@{}p{0.16\textwidth}p{0.29\textwidth}p{0.49\textwidth}@{}}
\toprule
\textbf{Node} & \textbf{Observed input in the realized trace} &
\textbf{Logged output summary} \\
\midrule
\multicolumn{3}{@{}l}{\textit{Original G-Designer execution}} \\
\texttt{6bRH} (Math Solver)
& Query only
& Computes 7 remaining rooms and returns $7\times 7=49$. \\
\texttt{4tCg} (Mathematical Analyst)
& Query only
& Computes total time $12\times7=84$, elapsed time $5\times7=35$, and returns $84-35=49$. \\
\texttt{3LK5} (Programming Expert)
& Query and the output of \texttt{4tCg}
& Generates a program that subtracts the painted rooms and returns 49. \\
\texttt{3Ssf} (Inspector)
& Query and the outputs of \texttt{4tCg} and \texttt{3LK5}
& Verifies the remaining-room calculation and returns 49. \\
FinalRefer
& Final outputs of all four agents
& Observes four consistent solutions and returns 49. \\
\midrule
\multicolumn{3}{@{}l}{\textit{E25N20 execution}} \\
\texttt{6bRH} (Math Solver)
& Query only
& Returns $7\times7=49$. \\
\texttt{3LK5} (Programming Expert)
& Query only
& Generates an equivalent program and returns 49. \\
\texttt{3Ssf} (Inspector)
& Query and the output of \texttt{3LK5}
& Checks the calculation and returns 49. \\
FinalRefer
& Final outputs of the three retained agents
& Receives three consistent answers and returns 49. \\
\bottomrule
\end{tabular}
\caption{Execution trace for the correct-to-correct MultiArith case.}
\label{tab:case_multiarith_419}
\end{table*}

The removed Mathematical Analyst is correct, but its derivation duplicates
information already available from the other roles and is forwarded to two
downstream agents. Removing this node shortens both the number of calls and
the prompts received downstream. Total online usage falls from 6,383 to
4,567 tokens, a reduction of $28.5\%$, while the answer remains correct.



\paragraph{Case 2: Wrong $\rightarrow$ Wrong under a shared domain misconception.}
The MMLU query asks which isolated pancreatic-enzyme loss would have the
most extensive effect on nutrient absorption in cystic fibrosis. The
reference option is C, trypsinogen. Both executions instead select A,
lipase. In the original trace,
$\texttt{4MEK}\!\rightarrow\!\texttt{4BNb}$,
$\texttt{4MEK}\!\rightarrow\!\texttt{4Cwg}$, and
$\texttt{4BNb}\!\rightarrow\!\texttt{4Cwg}$ are executed. E25N20 removes
\texttt{4MEK} (Knowledgeable Expert) and retains only
$\texttt{4BNb}\!\rightarrow\!\texttt{4Cwg}$.


\begin{table*}[t]
\centering
\scriptsize
\setlength{\tabcolsep}{4pt}
\renewcommand{\arraystretch}{1.12}
\begin{tabular}{@{}p{0.16\textwidth}p{0.29\textwidth}p{0.49\textwidth}@{}}
\toprule
\textbf{Node} & \textbf{Observed input in the realized trace} &
\textbf{Logged output summary} \\
\midrule
\multicolumn{3}{@{}l}{\textit{Original G-Designer execution}} \\
\texttt{4MEK} (Knowledgeable Expert)
& Query only
& Produces search terms covering cystic fibrosis, pancreatic function, lipase, trypsinogen, and nutrient absorption, without selecting an option. \\
\texttt{6wZy} (Critic)
& Query only
& Notes that trypsinogen activates proteolytic enzymes, but its response ends without a final option. \\
\texttt{4BNb} (Mathematician)
& Query and the output of \texttt{4MEK}
& Prioritizes fat and fat-soluble-vitamin malabsorption and selects A, lipase. \\
\texttt{3gkY} (Psychologist)
& Query only
& Uses the same fat-malabsorption argument and selects A. \\
\texttt{4Cwg} (Historian)
& Query and the outputs of \texttt{4MEK} and \texttt{4BNb}
& Adopts \texttt{4BNb}'s comparison and selects A. \\
FinalRefer
& Final outputs of all five agents
& Returns A, which disagrees with the reference option C. \\
\midrule
\multicolumn{3}{@{}l}{\textit{E25N20 execution}} \\
\texttt{6wZy} (Critic)
& Query only
& Again discusses the alternatives but does not produce a complete final option. \\
\texttt{4BNb} (Mathematician)
& Query only
& Again argues that lipase has the broadest nutritional consequences and selects A. \\
\texttt{3gkY} (Psychologist)
& Query only
& Selects A using the same fat-digestion rationale. \\
\texttt{4Cwg} (Historian)
& Query and the output of \texttt{4BNb}
& Repeats the lipase argument and selects A. \\
FinalRefer
& Final outputs of the four retained agents
& Returns A again. \\
\bottomrule
\end{tabular}
\caption{Execution trace for the wrong-to-wrong MMLU case.}
\label{tab:case_mmlu_62}
\end{table*}

This failure is not primarily caused by one removable noisy message. The
active reasoning roles share the same substantive misconception and rank the
direct consequences of lipase loss above the broader downstream role of
trypsinogen. Pruning reduces token usage from 6,314 to 5,550, but the retained
subgraph continues to support the same incorrect answer. This case exposes a
limit of topology refinement: removing redundant communication cannot repair
an error already shared by the retained agents.



\paragraph{Case 3: Correct $\rightarrow$ Wrong after removing a useful knowledge source.}
The MMLU query asks which worker uses a ``paddy wagon,'' with reference
option B, police officer. The original execution returns B. Its realized
trace is
$\texttt{4MEK}\!\rightarrow\!\texttt{4BNb}$,
$\texttt{4MEK}\!\rightarrow\!\texttt{4Cwg}$, and
$\texttt{4BNb}\!\rightarrow\!\texttt{4Cwg}$. E25N20 removes
\texttt{4MEK} (Knowledgeable Expert), retains only
$\texttt{4BNb}\!\rightarrow\!\texttt{4Cwg}$, and returns C, rice farmer.


\begin{table*}[t]
\centering
\scriptsize
\setlength{\tabcolsep}{4pt}
\renewcommand{\arraystretch}{1.12}
\begin{tabular}{@{}p{0.16\textwidth}p{0.29\textwidth}p{0.49\textwidth}@{}}
\toprule
\textbf{Node} & \textbf{Observed input in the realized trace} &
\textbf{Logged output summary} \\
\midrule
\multicolumn{3}{@{}l}{\textit{Original G-Designer execution}} \\
\texttt{4MEK} (Knowledgeable Expert)
& Query only
& Supplies lexical cues linking ``paddy wagon'' with police officer, while also listing the competing options. \\
\texttt{6wZy} (Critic)
& Query only
& Treats the wording as regionally ambiguous but identifies B as the most likely answer. \\
\texttt{4BNb} (Mathematician)
& Query and the output of \texttt{4MEK}
& Compares the police-van and agricultural readings and selects the common police-van meaning, B. \\
\texttt{3gkY} (Psychologist)
& Query only
& Independently interprets the term as a police transport vehicle and selects B. \\
\texttt{4Cwg} (Historian)
& Query and the outputs of \texttt{4MEK} and \texttt{4BNb}
& Follows the common slang interpretation and selects B. \\
FinalRefer
& Final outputs of all five agents
& Returns B, matching the reference answer. \\
\midrule
\multicolumn{3}{@{}l}{\textit{E25N20 execution}} \\
\texttt{6wZy} (Critic)
& Query only
& Still identifies B as the best answer, although with reservations about ambiguity. \\
\texttt{4BNb} (Mathematician)
& Query only
& Reinterprets the term literally as a wagon for transporting unmilled rice and selects C. \\
\texttt{3gkY} (Psychologist)
& Query only
& Independently retains the police-van interpretation and selects B. \\
\texttt{4Cwg} (Historian)
& Query and the output of \texttt{4BNb}
& Adopts the agricultural interpretation from \texttt{4BNb} and selects C. \\
FinalRefer
& Final outputs of the four retained agents
& Chooses C despite two independent agents selecting B. \\
\bottomrule
\end{tabular}
\caption{Execution trace for the correct-to-wrong MMLU case.}
\label{tab:case_mmlu_151}
\end{table*}

The removed node provides a useful lexical cue in the original run. After
its removal, the only retained communication path propagates an incorrect
literal interpretation from \texttt{4BNb} to \texttt{4Cwg}. The execution
uses $13.5\%$ fewer tokens, but the connected pair reinforces option C and
the final node rejects two independent correct responses. This example
shows that a lower-cost topology can lose complementary knowledge and
preserve a misleading information path.



\begin{figure*}[t]
    \centering
    \includegraphics[width=\textwidth]{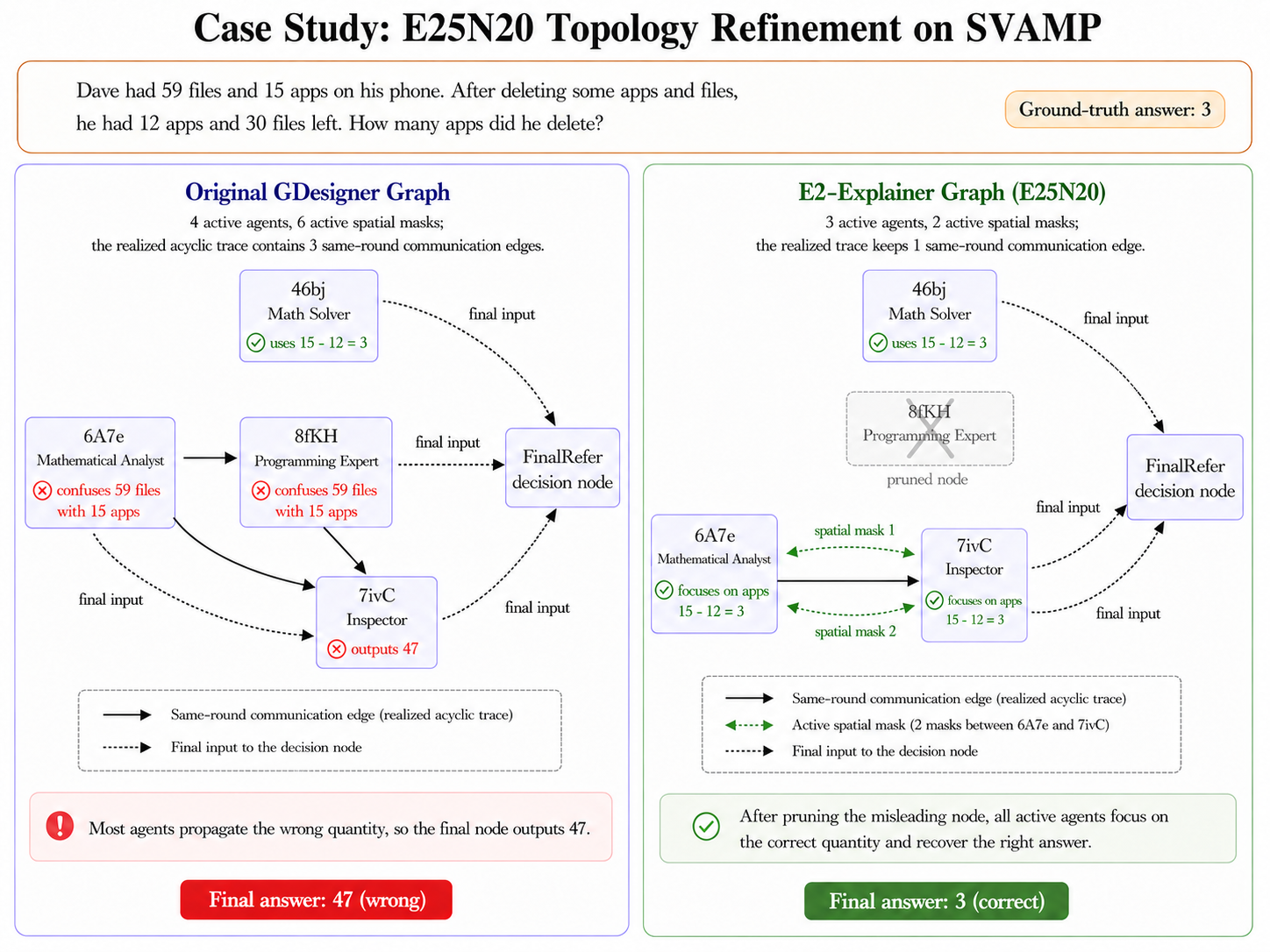}
    \caption{Wrong-to-correct case on SVAMP index 415. The original graph
    contains four active agents and six active spatial masks, and its
    realized acyclic trace contains three same-round communication edges.
    E25N20 removes the Programming Expert, retains three agents and two
    active spatial masks, and realizes one same-round communication edge.}
    \label{fig:case_study_svamp_415}
\end{figure*}

\paragraph{Case 4: Wrong $\rightarrow$ Correct after topology refinement.}
Figure~\ref{fig:case_study_svamp_415} visualizes the SVAMP example in which
Dave initially has 59 files and 15 apps, and retains 30 files and 12 apps
after deletion. The requested quantity is the number of deleted apps, so
the reference answer is $15-12=3$. The original G-Designer execution
returns 47, whereas E25N20 returns 3.


The original realized trace is
$\texttt{6A7e}\!\rightarrow\!\texttt{8fKH}$,
$\texttt{6A7e}\!\rightarrow\!\texttt{7ivC}$, and
$\texttt{8fKH}\!\rightarrow\!\texttt{7ivC}$. E25N20 removes
\texttt{8fKH} (Programming Expert), after which only
$\texttt{6A7e}\!\rightarrow\!\texttt{7ivC}$ is executed.

\begin{table*}[t]
\centering
\scriptsize
\setlength{\tabcolsep}{4pt}
\renewcommand{\arraystretch}{1.12}
\begin{tabular}{@{}p{0.16\textwidth}p{0.29\textwidth}p{0.49\textwidth}@{}}
\toprule
\textbf{Node} & \textbf{Observed input in the realized trace} &
\textbf{Logged output summary} \\
\midrule
\multicolumn{3}{@{}l}{\textit{Original G-Designer execution}} \\
\texttt{46bj} (Math Solver)
& Query only
& Correctly uses the app counts and returns $15-12=3$. \\
\texttt{6A7e} (Mathematical Analyst)
& Query only
& Misbinds 59 files as the original number of apps and returns $59-12=47$. \\
\texttt{8fKH} (Programming Expert)
& Query and the output of \texttt{6A7e}
& Sets \texttt{original\_apps=59} in its generated program and returns 47. \\
\texttt{7ivC} (Inspector)
& Query and the outputs of \texttt{6A7e} and \texttt{8fKH}
& Accepts the same variable assignment and returns 47. \\
FinalRefer
& Final outputs of all four agents
& Selects 47 from three mutually consistent but incorrect responses and rejects the isolated answer 3. \\
\midrule
\multicolumn{3}{@{}l}{\textit{E25N20 execution}} \\
\texttt{46bj} (Math Solver)
& Query only
& Returns $15-12=3$. \\
\texttt{6A7e} (Mathematical Analyst)
& Query only
& Correctly binds 59 and 30 to files and 15 and 12 to apps, then returns 3. \\
\texttt{7ivC} (Inspector)
& Query and the output of \texttt{6A7e}
& Verifies the app-based subtraction and returns 3. \\
FinalRefer
& Final outputs of the three retained agents
& Receives three consistent answers of 3 and returns 3. \\
\bottomrule
\end{tabular}
\caption{Execution trace for the wrong-to-correct SVAMP case in
Figure~\ref{fig:case_study_svamp_415}.}
\label{tab:case_svamp_415}
\end{table*}

In the original run, the incorrect variable binding is repeated along the
connected path and outweighs the only correct response at aggregation. In
the independent E25N20 run, all retained agents distinguish files from apps
and agree on 3. Token usage also falls from 6,059 to 4,965. The case
illustrates how a compact topology can coincide with both lower cost and a
more reliable aggregation outcome, while the independent sampling caveat
prevents attributing the correction solely to removal of \texttt{8fKH}.


\clearpage

\subsection{Generalization to Hand-Crafted Communication Topologies}
\label{app:handcrafted_topology}

We further examine whether the frozen explainer trained only on
G-Designer-generated graphs can transfer to manually specified
communication structures. We consider five representative topology
families, including complete, random, layered, chain, and star graphs.
E2-Explainer is directly applied to these graphs without additional
causal supervision or retraining.

\begin{strip}
\centering
\resizebox{\textwidth}{!}{
\begin{tabular}{llcccccc}
\toprule
Method
& Metric
& MMLU
& GSM8K
& MultiArith
& SVAMP
& AQuA
& HumanEval \\
\midrule

Complete
& Acc.
& 77.39
& 90.71
& 97.50
& 94.31
& 83.90
& 87.09
\\

\multirow{2}{*}{Complete + E2-Explainer}
& Acc.
& 78.43{\scriptsize $\uparrow$1.04}
& 91.35{\scriptsize $\uparrow$0.64}
& 97.67{\scriptsize $\uparrow$0.17}
& 94.50{\scriptsize $\uparrow$0.19}
& 84.64{\scriptsize $\uparrow$0.74}
& 88.70{\scriptsize $\uparrow$1.61}
\\
& Token $\Delta$
& -17.4\%
& -36.5\%
& -30.9\%
& -22.4\%
& -28.9\%
& -25.1\%
\\

\midrule

Random
& Acc.
& 76.78
& 91.01
& 97.17
& 93.97
& 84.70
& 86.88
\\

\multirow{2}{*}{Random + E2-Explainer}
& Acc.
& 77.45{\scriptsize $\uparrow$0.67}
& 90.97{\scriptsize $\downarrow$0.04}
& 97.33{\scriptsize $\uparrow$0.16}
& 94.09{\scriptsize $\uparrow$0.12}
& 85.03{\scriptsize $\uparrow$0.33}
& 87.09{\scriptsize $\uparrow$0.21}
\\
& Token $\Delta$
& -18.7\%
& -21.2\%
& -19.7\%
& -15.6\%
& -17.6\%
& -14.3\%
\\

\midrule

Layer
& Acc.
& 77.08
& 91.24
& 96.83
& 94.40
& 85.49
& 88.38
\\

\multirow{2}{*}{Layer + E2-Explainer}
& Acc.
& 78.10{\scriptsize $\uparrow$1.02}
& 91.28{\scriptsize $\uparrow$0.04}
& 97.41{\scriptsize $\uparrow$0.58}
& 94.91{\scriptsize $\uparrow$0.51}
& 85.43{\scriptsize $\downarrow$0.06}
& 89.11{\scriptsize $\uparrow$0.73}
\\
& Token $\Delta$
& -23.6\%
& -30.1\%
& -27.4\%
& -22.8\%
& -33.9\%
& -19.9\%
\\

\midrule

Chain
& Acc.
& 78.05
& 91.09
& 96.67
& 94.62
& 83.51
& 86.29
\\

\multirow{2}{*}{Chain + E2-Explainer}
& Acc.
& 78.37{\scriptsize $\uparrow$0.32}
& 90.67{\scriptsize $\downarrow$0.42}
& 96.67{\scriptsize $\pm$0.00}
& 94.95{\scriptsize $\uparrow$0.33}
& 83.46{\scriptsize $\downarrow$0.05}
& 85.88{\scriptsize $\downarrow$0.41}
\\
& Token $\Delta$
& -15.2\%
& -25.0\%
& -24.1\%
& -14.3\%
& -15.4\%
& -7.93\%
\\

\midrule

Star
& Acc.
& 77.08
& 90.78
& 97.67
& 93.85
& 83.90
& 83.25
\\

\multirow{2}{*}{Star + E2-Explainer}
& Acc.
& 78.03{\scriptsize $\uparrow$0.95}
& 91.43{\scriptsize $\uparrow$0.65}
& 97.83{\scriptsize $\uparrow$0.16}
& 94.17{\scriptsize $\uparrow$0.32}
& 85.82{\scriptsize $\uparrow$1.92}
& 86.13{\scriptsize $\uparrow$2.88}
\\
& Token $\Delta$
& -29.6\%
& -35.4\%
& -22.8\%
& -17.9\%
& -32.5\%
& -20.1\%
\\

\bottomrule
\end{tabular}
}
\captionof{table}{
Generalization of E2-Explainer to manually specified communication
topologies. The explainer is trained only on G-Designer-generated
graphs and applied without retraining. Accuracy arrows indicate changes
relative to the corresponding unrefined topology.
}
\label{tab:handcrafted_topologies}
\end{strip}

Table~\ref{tab:handcrafted_topologies} shows that E2-Explainer
transfers effectively beyond the generated graph distribution used for
training. It improves or matches accuracy in 26 of the 30
dataset--topology combinations while reducing token usage across every
topology and benchmark. The strongest gains appear on the star topology,
including improvements of 1.92 points on AQuA and 2.88 points on
HumanEval. The layered topology also benefits consistently, reaching
89.11 on HumanEval while reducing token usage by 19.9\%. Even for the
chain topology, whose sparse structure leaves less removable redundancy,
the explainer largely preserves task performance and still reduces token
usage on all six benchmarks.

These results are notable because the explainer is calibrated exclusively
on G-Designer graphs but is directly applied to fixed topologies with
substantially different structural patterns. Its improvements therefore
cannot be attributed solely to memorizing G-Designer-specific adjacency
structures. Instead, the explainer captures transferable task- and
structure-dependent cues for retaining useful communication elements.

\subsection{Additional Redundancy Analysis on OFA-MAS}
\label{app:ofa_random_pruning}

\begin{figure*}[!t]
    \centering
    \includegraphics[width=0.90\textwidth]{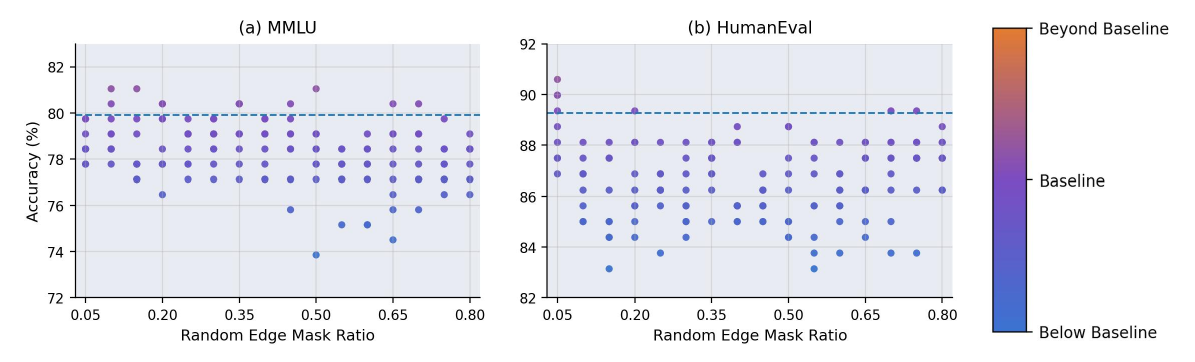}
    \caption{Random edge masking on OFA-MAS-generated
    communication graphs for MMLU and HumanEval. The edge
    masking ratio ranges from 0.05 to 0.80 in increments of
    0.05. Each point denotes the accuracy obtained under one
    retained random mask, while the dashed line marks the
    performance of the original unmasked OFA-MAS graph.
    For visualization, one extreme run is omitted at each
    masking ratio. The resulting distribution shows that although a few random
    masks preserve or even improve the original performance,
    most random masks degrade accuracy, indicating that
    task-preserving compact subgraphs exist but are difficult
    to identify through unguided pruning.}
    \label{fig:ofamas_random_pruning}
\end{figure*}

To examine whether the redundancy observed on
G-Designer is specific to a particular topology designer,
we repeat the same random edge-masking analysis on
OFA-MAS, as shown in Figure~\ref{fig:ofamas_random_pruning}.
Similar to the observations on G-Designer, most random edge
masks degrade performance on both MMLU and HumanEval,
while only a small fraction of them preserve or slightly
improve the original accuracy. This pattern suggests that
OFA-MAS-generated topologies also contain redundant
communication links and that task-preserving compact
subgraphs do exist.

At the same time, the rarity of these successful random
masks shows that such subgraphs are difficult to identify
through unguided pruning alone. In other words, redundancy
does not imply that arbitrary pruning is safe. This further
motivates the need for a task-conditioned explainer that
selects compact communication subgraphs according to causal
contribution rather than random deletion.

\FloatBarrier


\section{Algorithmic Details}
\label{app:algorithms}

\subsection{Causal Supervision Construction and Explainer Prediction}
Algorithms~\ref{alg:causal_dataset} and~\ref{alg:amortized_explainer}
summarize the offline construction of graph-to-subgraph supervision
and the training and test-time prediction procedures, respectively.

\begin{algorithm}[H]
\footnotesize
\caption{Causally Grounded Supervision Construction}
\label{alg:causal_dataset}
\begin{algorithmic}[1]
\Require $\mathcal{D}_{\mathrm{cal}},\mathcal{G}_{\phi},
\mathcal{B},M,\alpha,\beta$
\Ensure Graph-to-subgraph dataset $\mathcal{D}_{\mathrm{expl}}$
\State $\mathcal{D}_{\mathrm{expl}}\gets\emptyset$
\For{each $x\in\mathcal{D}_{\mathrm{cal}}$}
    \State $G=(V,E)\gets\mathcal{G}_{\phi}(x)$
    \For{each active edge $e\in E$}
        \State $(Q_G^{e},\mathcal{Y}^{G,e})
        \gets\Call{Evaluate}{x,G,M}$
        \State $G^{-e}\gets\Call{Intervene}{G,e}$
        \State $(Q_e,\mathcal{Y}^{G^{-e}})
        \gets\Call{Evaluate}{x,G^{-e},M}$
        \State $u^{\mathrm{task}}_e
        \gets\max(0,Q_G^{e}-Q_e)$
        \State $u^{\mathrm{sem}}_e
        \gets\max\!\left(
        0,\bar{H}(\mathcal{Y}^{G^{-e}})
        -\bar{H}(\mathcal{Y}^{G,e})
        \right)$
        \State $\widetilde{u}^{\mathrm{task}}_e
        \gets u^{\mathrm{task}}_e$
        \State $\widetilde{u}^{\mathrm{sem}}_e
        \gets\min(1,u^{\mathrm{sem}}_e/\log M)$
        \State $u_e\gets
        \operatorname{clip}_{[0,1]}
        (\alpha\widetilde{u}^{\mathrm{task}}_e+
        \beta\widetilde{u}^{\mathrm{sem}}_e)$
    \EndFor
    \State $\pi_G\gets\Call{SortDescending}{E,\{u_e\}_{e\in E}}$
    \For{each budget $b\in\mathcal{B}$}
        \State $(k_E,k_V)\gets\Call{RetentionCounts}{G,b}$
        \State $E_b^{\star}\gets\Call{TopK}{\pi_G,k_E}$
        \State $H_b^{\star}\gets
        \mathcal{R}_{\mathrm{valid}}(G,E_b^{\star},k_V)$
        \State $\mathcal{D}_{\mathrm{expl}}\gets
        \mathcal{D}_{\mathrm{expl}}
        \cup\{(x,G,b,H_b^{\star})\}$
    \EndFor
\EndFor
\State \Return $\mathcal{D}_{\mathrm{expl}}$
\end{algorithmic}
\end{algorithm}

\begin{algorithm}[H]
\footnotesize
\caption{Explainer Training and Test-Time Prediction}
\label{alg:amortized_explainer}
\begin{algorithmic}[1]
\Require $\mathcal{D}_{\mathrm{expl}},
\lambda_{\mathrm{node}},\lambda_{\mathrm{wd}}$
\Ensure Trained explainer $F_{\theta}$
\State Initialize parameters $\theta$
\Statex \textbf{Training}
\For{each minibatch $\mathcal{S}\subseteq
\mathcal{D}_{\mathrm{expl}}$}
    \For{each $(x_i,G_i,b,H_{i,b}^{\star})\in\mathcal{S}$}
        \State $(\mathbf{A}_{i,b}^{\star},
        \mathbf{R}_{i,b}^{\star})
        \gets\Call{Indicators}{G_i,H_{i,b}^{\star}}$
        \State $(\widehat{\mathbf{A}}_{i,b},
        \widehat{\mathbf{R}}_{i,b})
        \gets F_{\theta}(x_i,G_i,b)$
    \EndFor
    \State Compute $\mathcal{L}_{\mathrm{edge}}$
    and $\mathcal{L}_{\mathrm{node}}$ on $\mathcal{S}$
    \State $\mathcal{L}\gets
    \mathcal{L}_{\mathrm{edge}}
    +\lambda_{\mathrm{node}}\mathcal{L}_{\mathrm{node}}
    +\lambda_{\mathrm{wd}}\lVert\theta\rVert_2^2$
    \State Update $\theta$ by minimizing $\mathcal{L}$
\EndFor
\Statex \textbf{Test-time prediction}
\State Receive query $x$, graph $G$, and budget $b$
\State $(\widehat{\mathbf{A}}_b,
\widehat{\mathbf{R}}_b)\gets F_{\theta}(x,G,b)$
\State $(k_E,k_V)\gets\Call{RetentionCounts}{G,b}$
\State $\widehat{E}_b\gets
\Call{TopK}{\widehat{\mathbf{A}}_b,k_E}$
\State $\widehat{V}_b\gets
\Call{TopK}{\widehat{\mathbf{R}}_b,k_V}$
\State $\widehat{H}_b\gets
\mathcal{R}_{\mathrm{valid}}
(G,\widehat{E}_b,\widehat{V}_b)$
\State \Return $\widehat{H}_b$
\end{algorithmic}
\end{algorithm}
